\documentclass{article}

\PassOptionsToPackage{numbers, compress}{natbib}
\usepackage[dblblindworkshop, final]{neurips_2026}

\workshoptitle{The 6th Workshop on Mathematical Reasoning and AI}

\usepackage[utf8]{inputenc} % allow utf-8 input
\usepackage[T1]{fontenc}    % use 8-bit T1 fonts
\usepackage{hyperref}       % hyperlinks
\usepackage{url}            % simple URL typesetting
\usepackage{booktabs}       % professional-quality tables
\usepackage{amsfonts}       % blackboard math symbols
\usepackage{nicefrac}       % compact symbols for 1/2, etc.
\usepackage{microtype}      % microtypography
\usepackage{xcolor}         % colors
\usepackage{graphicx}
\usepackage{float}
\usepackage{fvextra}

\definecolor{codegreen}{rgb}{0,0.6,0}
\definecolor{codegray}{rgb}{0.5,0.5,0.5}
\definecolor{codepurple}{rgb}{0.58,0,0.82}
\definecolor{backcolour}{rgb}{0.95,0.95,0.92}
\definecolor{darkgreen}{rgb}{0.0,0.4,0.0}
\definecolor{internalblue}{RGB}{40,90,160}

\newcommand{\para}[1]{\noindent\textbf{#1}\hspace{0.5em}\ignorespaces}

\usepackage{listings}
\definecolor{panelbg}{gray}{0.97}
\definecolor{panelrule}{RGB}{110,120,135}
\newcommand{\panellabel}[1]{{\scriptsize\sffamily\bfseries\color{panelrule}#1}}
\lstdefinestyle{panelbase}{
  basicstyle=\scriptsize\ttfamily,
  backgroundcolor=\color{panelbg},
  frame=tb, rulecolor=\color{panelrule}, framerule=0.5pt, framesep=5pt,
  breaklines=true, breakindent=6pt, columns=fullflexible, keepspaces=true,
  showstringspaces=false,
  aboveskip=10pt, belowskip=10pt
}
\lstnewenvironment{jsonpanel}[1]{\lstset{style=panelbase,
  morestring=[b]", stringstyle=\color{internalblue},
  title={\panellabel{#1}}}}{}
\newcommand{\phead}[1]{{\ttfamily\bfseries\color{internalblue}#1}}
\lstnewenvironment{promptpanel}[1]{\lstset{style=panelbase,
  title={\panellabel{#1}},
  escapeinside={(*@}{@*)}}}{}

\newcommand{\rev}[1]{{\color{black}#1}}

\title{What Has Actually Been Proved? Auditable Paper-to-Lean Alignment with LeanTrail}

\author{%
  Fabricio Venturim \\
  Fundação Getulio Vargas \\
  Rio de Janeiro, Brazil \\
  \texttt{fabricio.venturim@fgv.br} \\
  \And
  Leighton Estrada \\
  Fundação Getulio Vargas \\
  Rio de Janeiro, Brazil \\
  \texttt{leighton.rayme@fgv.br} \\
  \AND
  Claudio Silva \\
  New York University \\
  New York, USA \\
  \texttt{csilva@nyu.edu} \\
  \And
  Jorge Poco \\
  Fundação Getulio Vargas \\
  Rio de Janeiro, Brazil \\
  \texttt{jorge.poco@fgv.br} \\
}

\begin{document}

\maketitle

\begin{abstract}
Compiler feedback certifies that Lean accepts a declaration, not that the declaration captures the intended result: a repository can compile while a central theorem is weakened, merely auxiliary, or transitively dependent on \texttt{sorry}. Formalization progress should instead be measured against the source paper, \rev{in terms of its \emph{paper items}}: definitions, lemmas, and theorems. \textit{LeanTrail} implements this view as a post-hoc, tool-agnostic audit: deterministic retrieval proposes candidate counterparts for each paper item in the Lean repository and, when available, the Blueprint; a language model judges statement-level correspondence, only among those candidates; and proof completeness comes from a deterministic transitive dependency audit never exposed to the model. Fusion assigns each item one of five coverage states, from \textit{Formalized} to \textit{Missing}, each backed by inspectable evidence. On consensus annotations of 98 items from three Lean projects, \rev{with three runs of each of two judge models, \textit{LeanTrail} assigns the correct coverage state to 91.8\% (Gemini~3.8~Flash) and 92.5\% (GPT-6~Sol) of items. 96--97 of the 98 states are identical across runs, and GPT-6~Sol never reports a \textit{Partial} item as \textit{Formalized}.} Uncovered items form a natural task queue for formalization agents, anchored to the specification rather than the compiler.
\end{abstract}

\section{Introduction}
\looseness=-1
Formalization projects in proof assistants such as Lean~\cite{demoura2021lean4}, supported by libraries such as \textit{mathlib}~\cite{mathlib2020}, have become a practical medium for verifying research mathematics. Large language models increasingly translate informal statements into formal ones~\cite{wu2022autoformalization}, guide proof search~\cite{jiang2023dsp}, and act as agents that iterate against the compiler~\cite{yang2023leandojo}. As producing compilable Lean code becomes easier, a second bottleneck emerges: determining whether the intended mathematics has actually been formalized.

Compiler feedback certifies that Lean accepts a declaration, not that the declaration is faithful to the source paper: a repository can compile while a central theorem depends transitively on \texttt{sorry}, or while a declaration proves something weaker or merely auxiliary. Checking progress therefore requires reasoning across artifacts: Is Theorem~2 actually proved? Does its Lean statement match the paper's? These judgments matter in human--AI workflows, where mathematicians delegate mechanical work but keep mathematical control~\cite{collins2026workflows}.

Existing systems address parts of this workflow. \textit{Lea}~\cite{lea2026} supports mathematician-led formalization and audits generated Lean artifacts; Blueprints~\cite{massot2020leanblueprint,tao2023pfr} and \textit{LeanArchitect}~\cite{zhu2026leanarchitect} connect a formalization plan to formal declarations; \textit{Lean Atlas}~\cite{leanatlas2026} scopes human semantic review over the Lean dependency graph, and \textit{KnowTeX}~\cite{knowtex2026} exposes dependencies within the \LaTeX{} source. All are organized around declarations or plan nodes, and none answers the question a project targeting an existing paper raises: which of the paper's results are formalized? Declarations do not say which paper items they cover, and a Blueprint is a re-authored plan that can drift from the paper it describes. \rev{Autoformalization metrics instead judge a given pair: BEq+~\cite{poiroux2025beq} compares a generated formal statement with a reference formal one, and BPF~\cite{mohammad2026faithfulness} scores how faithfully a formal candidate encodes a natural-language statement by probing its consequences. \textit{LeanTrail} must first find the candidate among a repository's declarations and combine its judgment with the proof audit; a certifier such as BPF could replace or check that judgment.} We argue that formalization progress should be measured against the paper itself, \rev{in terms of its paper items}: definitions, lemmas, propositions, and theorems.

\textit{LeanTrail} answers \rev{that question} with a post-hoc audit of any paper and repository (Figure~\ref{fig:pipeline}), built to keep apart two questions that progress reports usually conflate. Whether a Lean statement says what the paper says is a semantic question; a language model answers it, but only over a short list of deterministically retrieved candidates. Whether that statement is proved is a mechanical one; we answer it by walking the elaborated dependency graph for \texttt{sorryAx}, and the model never sees the result. Fixed rules fuse the two answers into one of five coverage states per item, each shown with the candidates it rejected. The split has a concrete consequence: a model error can mislabel how faithful a statement is, but cannot turn an unproved result into a proved one. Five states rather than a percentage then tell a contributor what to do next.

Our contributions are: (i) a paper-level formulation of formalization progress that is auditable by construction, since the model proposes and classifies but never certifies a proof; (ii) a pipeline that keeps deterministic retrieval, model-judged correspondence, and deterministic proof auditing separate; (iii) an interface for inspecting and contesting paper-to-Lean coverage; and (iv) an evaluation on 98 consensus-annotated items from three projects, \rev{with two judge models and three runs each: \textit{LeanTrail} recovers the correct state for 91.8--92.5\% of items and repeated runs agree on 96--97 of 98 states}, with the annotations released.
\section{Alignment and Coverage Classification}
\label{sec:alignment}
\begin{figure}[t]
  \centering
  \includegraphics[width=0.80\textwidth]{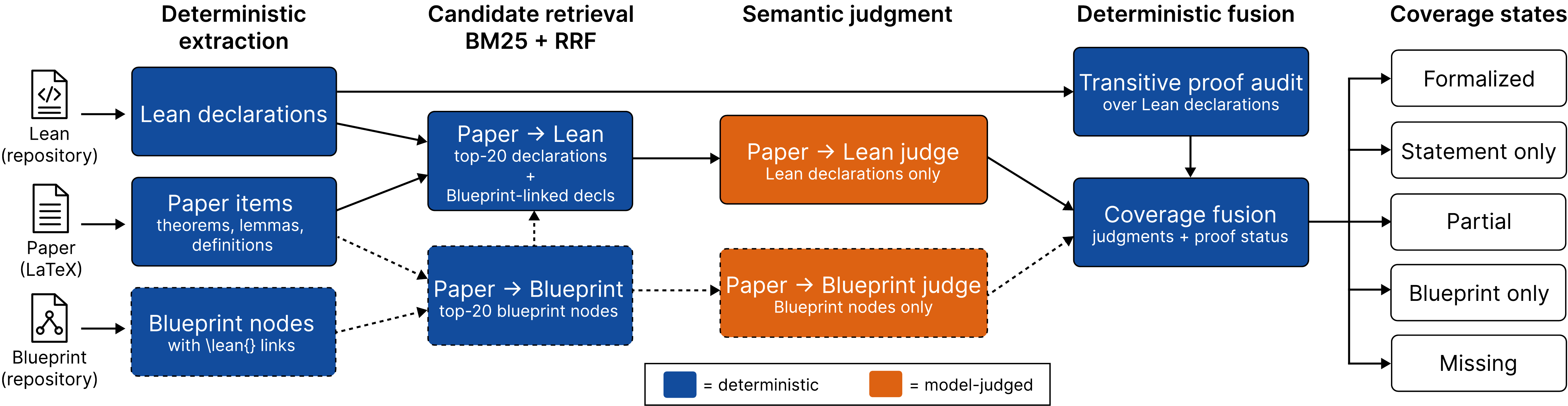}
  \caption{The \textit{LeanTrail} pipeline. Blue stages are deterministic: extraction, candidate retrieval, the transitive proof audit, and the final fusion. The only model-judged stages (orange) label the statement-level semantic relation of retrieved candidates; dashed elements require a Blueprint.}
  \label{fig:pipeline}
\end{figure}

\looseness=-1
\textit{LeanTrail} converts a formalization project into a precomputed paper-level representation: for each item in the paper, it records candidate Lean declarations and, when available, Blueprint nodes, with one coverage state. Figure~\ref{fig:pipeline} summarizes the pipeline: deterministic stages (blue) extract entities, retrieve candidates, audit proof completeness, and fuse the results, while the only model-judged stages (orange) classify semantic correspondence and can never certify that a proof is complete.

\para{Extraction.}
From the paper, \textit{LeanTrail} extracts theorem-like \LaTeX{} environments (theorems, lemmas, propositions, definitions, including paper-defined variants), recording each item's kind, label, statement, and source location. From the Lean repository, it indexes declarations, signatures, source ranges, and the constants referenced by each elaborated declaration, which form the declaration-dependency graph used by the proof audit. From the optional Blueprint, it extracts nodes, statements, dependencies, and explicit links to Lean declarations (details in Appendix~\ref{app:pipeline}).

\para{Candidate retrieval.}
Rather than asking the model to examine an entire repository, \textit{LeanTrail} narrows each search space deterministically, as in retrieval-augmented generation~\citep{lewis2020retrieval} and premise selection~\citep{yang2023leandojo}. Alignment follows two routes, Paper$\rightarrow$Lean over declarations and Paper$\rightarrow$Blueprint over plan nodes; each fuses BM25 textual rankings~\citep{robertson2009probabilistic} with identifier similarity via reciprocal rank fusion~\citep{cormack2009reciprocal} and retains the top 20 candidates\rev{, the smallest cutoff we tested at which every annotated counterpart is retrieved on both routes (Table~\ref{tab:alignment-results}). This is an empirical choice for these projects, not a guarantee for larger repositories, where the cutoff can be raised}. For each retrieved Blueprint node with an explicit Lean link, the linked declaration is added to the Lean candidate pool; these links improve recall, but their provenance is hidden from the Lean judge.

\begin{figure}[t]
  \centering
  \includegraphics[width=0.86\textwidth]{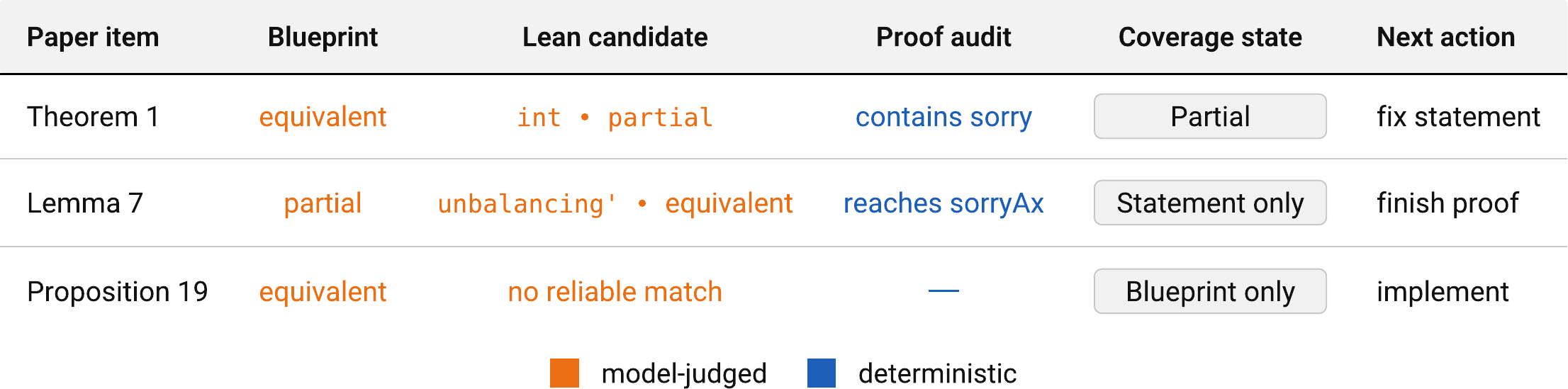}
  \caption{Three APAP paper items and the evidence behind their coverage states \rev{in a Gemini~3.8~Flash run}, as read from the interface (Figure~\ref{fig:paper-alignment}, Appendix~\ref{app:views}): semantic relations are model-judged (orange), the proof audit deterministic (blue).}
  \label{fig:coverage-strip}
\end{figure}

\para{Semantic judgment.}
Two independent model calls, one per route, classify the retained candidates as \textit{Equivalent}, \textit{Partial}, or \textit{None}. \textit{Equivalent} denotes the same substantive result, allowing representation changes or direct instantiation. \textit{Partial} denotes the same intended result with a material mismatch in hypotheses, domain, parameters, or conclusion. \textit{None} covers proof ingredients, related results, and unrelated results. The model receives statement-level information only: declaration bodies, proof-completeness data, and cross-route evidence are withheld, and it neither searches the repository nor decides whether a proof is complete. The prompt policy, response schema, and exact templates are given in Appendix~\ref{app:prompts}.

\para{Proof audit.}
In parallel with semantic judgment, \textit{LeanTrail} propagates \texttt{sorryAx} deterministically through the elaborated declaration-dependency graph, so a declaration counts as incomplete even when a \texttt{sorry} enters only through its dependencies. If a declaration's status cannot be resolved from the graph alone, a targeted \texttt{\#print axioms} certificate settles it before fusion. \rev{Besides \texttt{sorry}, a proof can rest on an \emph{axiom}, a statement that a project asserts without proof and that Lean accepts just as silently. The certificate names every axiom a declaration depends on; in the evaluated projects, no project declares an axiom of its own, and every certificate lists only the three axioms of Lean's standard foundations (\texttt{propext}, \texttt{Classical.choice}, \texttt{Quot.sound}), which \textit{mathlib} itself assumes.} \rev{The model never sees the outcome of this audit, so whether a proof is complete cannot sway its judgment of whether a statement matches the paper.}

\para{Fusion into coverage states.}
The final state combines semantic correspondence with proof status; a \emph{local} match is a declaration in the project repository itself. \textit{Formalized} requires an equivalent local declaration whose proof (or implementation, for definition-like items) is complete; \textit{Statement only}, an equivalent Lean proposition whose audit is incomplete or unknown; \textit{Partial}, a partial relation on either route with no equivalent local match; \textit{Blueprint only}, an equivalent Blueprint node without a validated local match; and \textit{Missing}, no equivalent or partial match on either route. Applied in this order, the rules give each item exactly one state.
\section{Inspecting the Alignment}
\label{sec:inspecting}

Coverage states are only useful if a user can check them, so the interface makes each decision inspectable and open to challenge. From a paper item, a user reaches its Lean counterpart and proof status, the retained candidates with their judged relations and rationales, and a coverage table over all items, so a questionable state can be traced to the judgment or audit that produced it, including the rejected alternatives. \rev{A \textit{Formalized} state therefore still needs checking: the proof is verified mechanically, but whether the Lean statement matches the paper is the model's judgment.} Figure~\ref{fig:paper-alignment} (Appendix~\ref{app:views}) shows these views on the example below.

Figure~\ref{fig:coverage-strip} shows why the five states are worth separating, on APAP (Arithmetic Progressions and Almost Periodicity)~\cite{bloom2023,dillies2026apap}, a formalization of Kelley--Meka-type bounds for sets free of three-term arithmetic progressions. All three items have a Blueprint counterpart (equivalent for Theorem~1 and Proposition~19, partial for Lemma~7), yet end in different states and call for different next actions. The differences are substantive: \texttt{int} quantifies $c>0$ after $A$ and $N$, allowing the constant to depend on them, whereas the paper requires an absolute constant; \texttt{unbalancing'} states \textit{Lemma~7} faithfully, but its transitive dependency closure reaches \texttt{sorryAx}. No single progress number distinguishes these cases, and \S\ref{sec:reliability} measures whether \textit{LeanTrail} recovers the distinction reliably.

\vspace{-0.1in}
\section{How Reliable Is the Alignment?}
\label{sec:reliability}

\begin{table}[t]
\centering
\caption{Performance across the three evaluated stages of \textit{LeanTrail} (\%). $N_P$ and $N_L$ denote paper items and indexed Lean declarations, and $N_+$ the positive queries per route\rev{; @$k$ is query-level recall at cutoff $k$}. \rev{Judgment and coverage accuracies are means over three runs of each judge model, with the standard deviation in the overall row.}}
\label{tab:alignment-results}
\small
\textbf{(a)} Candidate retrieval\par\smallskip
{\setlength{\tabcolsep}{3pt}%
\begin{tabular}{lcccccccccccccc}
\toprule
& & & \multicolumn{6}{c}{Paper$\rightarrow$Lean} & \multicolumn{6}{c}{Paper$\rightarrow$Blueprint} \\
\cmidrule(lr){4-9}\cmidrule(lr){10-15}
\textbf{Project} & $N_P$ & $N_L$ & $N_+$ & \textbf{@1} & \textbf{@5} & \textbf{@10} & \textbf{@15} & \textbf{@20} & $N_+$ & \textbf{@1} & \textbf{@5} & \textbf{@10} & \textbf{@15} & \textbf{@20} \\
\midrule
FinEqs & 7 & 11 & 6 & 50.0 & 100.0 & 100.0 & 100.0 & 100.0 & 7 & 100.0 & 100.0 & 100.0 & 100.0 & 100.0 \\
APAP & 29 & 929 & 12 & 50.0 & 75.0 & 83.3 & 91.7 & 100.0 & 24 & 83.3 & 95.8 & 100.0 & 100.0 & 100.0 \\
Ringel & 62 & 396 & 17 & 41.2 & 70.6 & 82.4 & 94.1 & 100.0 & 6 & 50.0 & 83.3 & 83.3 & 100.0 & 100.0 \\
\midrule
\textbf{Overall} & 98 & 1,336 & 35 & 45.7 & 77.1 & 85.7 & 94.3 & 100.0 & 37 & 81.1 & 94.6 & 97.3 & 100.0 & 100.0 \\
\bottomrule
\end{tabular}}

\medskip
\textbf{(b)} Semantic judgment and coverage\par\smallskip
{\setlength{\tabcolsep}{4pt}%
\begin{tabular}{lcccccc}
\toprule
& \multicolumn{3}{c}{\textbf{Gemini~3.8~Flash}} & \multicolumn{3}{c}{\textbf{GPT-6~Sol}} \\
\cmidrule(lr){2-4}\cmidrule(lr){5-7}
& \multicolumn{2}{c}{Judgment} & Coverage & \multicolumn{2}{c}{Judgment} & Coverage \\
\textbf{Project} & Paper$\rightarrow$Lean & Paper$\rightarrow$Blueprint & 5-state & Paper$\rightarrow$Lean & Paper$\rightarrow$Blueprint & 5-state \\
\midrule
FinEqs & 100.0 & 100.0 & 100.0 & 85.7 & 100.0 & 85.7 \\
APAP & 87.4 & 86.2 & 87.4 & 82.8 & 79.3 & 82.8 \\
Ringel & 93.0 & 96.8 & 93.0 & 97.8 & 93.5 & 97.8 \\
\midrule
\textbf{Overall} & 91.8$\pm$1.0 & 93.9$\pm$0.0 & 91.8$\pm$1.0 & 92.5$\pm$0.6 & 89.8$\pm$0.0 & 92.5$\pm$0.6 \\
\bottomrule
\end{tabular}}
\end{table}

\para{Evaluation setup.}
To measure how often assigned states are wrong, and in which direction, we evaluate all three stages, retrieval, semantic judgment, and coverage classification, against an item-level ground truth at fixed commits over three paper--repository pairs: FinEqs~\cite{baranczuk2019fineqs,naskrecki2025fineqs}, on reducing the number of equations defining a subset of $n$-space over a finite field (7 items, 11 indexed Lean declarations); APAP~\cite{bloom2023,dillies2026apap} (29 items, 929 declarations); and Ringel~\cite{montgomery2021ringel,elkersh2026ringel}, on decomposing complete graphs into copies of a fixed tree (62 items, 396 declarations). All 98 items come from explicit statement environments, checked against the source papers before annotation.

Two annotators familiar with Lean jointly reviewed every item against the repository snapshot and, when available, the Blueprint and project mappings, recording by consensus the corresponding Lean declaration, if any, its relation to the paper statement, and the coverage state. Project links were treated as evidence rather than labels, and proof completeness came from the transitive audit of \S\ref{sec:alignment}, not from manual judgment. The ground truth comprises 7 \textit{Formalized}, 5 \textit{Statement only}, 23 \textit{Partial}, 13 \textit{Blueprint only}, and 50 \textit{Missing} items (protocol and rationales in Appendix~\ref{app:annotation}, released with the supplementary material).

Table~\ref{tab:alignment-results} reports each stage: query-level retrieval recall@$k$, where a query is positive when the ground truth names a counterpart on that route; semantic-judgment accuracy over \textit{Equivalent}, \textit{Partial}, and \textit{None}; and five-state coverage accuracy. \rev{We run the semantic judgment with Gemini~3.8~Flash and GPT-6~Sol, three times each on identical inputs, and report means over the runs (configuration in Appendix~\ref{app:prompts}).} Appendix~\ref{app:evaluation} defines the metrics\rev{.}

\para{Results.}\looseness=-1
At $k=20$, retrieval recovered an annotated counterpart for every positive query on both routes, although recall at $k=1$ is low (45.7\% and 81.1\%), so candidates reach the judge as a pool. \rev{Paper$\rightarrow$Lean judgment reached 91.8\% accuracy (84.1\% macro-F1) with Gemini~3.8~Flash and 92.5\% (80.7\%) with GPT-6~Sol, and Paper$\rightarrow$Blueprint 93.9\% (84.7\%) and 89.8\% (77.7\%). After fusion with the proof audit, \textit{LeanTrail} assigned the correct coverage state to 91.8\% of items (five-state macro-F1 83.9\%) with Gemini and 92.5\% (77.6\%) with GPT-6~Sol, and separating any coverage from \textit{Missing} was correct for at least 97 of 98 items in every run. Repeated runs agree on 96 (Gemini) and 97 (GPT-6~Sol) of the 98 states, so the remaining errors are systematic rather than sampling noise.}

\para{Failure modes.}\looseness=-1
\rev{We separate errors by direction. An error \emph{over-claims} when it reports more coverage than annotated, such as a \textit{Partial} item as \textit{Formalized} or \textit{Statement only}, and \emph{under-claims} when it reports less. Over-claiming is the harmful direction for an audit, since a user may then trust an unfaithful statement. Gemini~3.8~Flash over-claims on 4.4\% of items per run and under-claims on 3.7\%. Of its \textit{Formalized} reports, 17.9\% are wrong. GPT-6~Sol never over-claims: every \textit{Formalized} report is correct and every \textit{Partial} item detected. All its errors, 7.5\% of items per run, under-claim faithful items as \textit{Partial} for differences the annotators judged immaterial, such as a rounded threshold. As a result, it recognizes 52.4\% of \textit{Formalized} items, against 85.7\% for Gemini. The models thus differ in the direction of their errors rather than in accuracy: GPT-6~Sol leads on Ringel, whose partial matches hide subtle mismatches, and trails on APAP and FinEqs, whose faithful matches carry harmless restrictions.}

\vspace{-0.1in}

\section{Toward an Agentic Formalization Loop}
\label{sec:loop}

\looseness=-1
\textit{LeanTrail} turns every non-\textit{Formalized} paper item into a structured, agent-ready task; the required action follows from the coverage state: \textit{Missing} calls for a new formalization, \textit{Blueprint only} for implementing a planned result, \textit{Statement only} for completing a proof, and \textit{Partial} for repairing a mismatch or accepting an intentional deviation. \rev{Each task bundles the evidence behind its state with a completion criterion and is exported as JSON for downstream agents (Appendix~\ref{app:agent-tasks}).}

\looseness=-1
Rerunning the audit after repository changes\rev{, such as an agent's edits,} recomputes each task against the same specification. \rev{Judgments are reused while the paper statement and the chosen counterparts are unchanged, while candidates and proof status are recomputed on every rerun.} Discharging the last transitive \texttt{sorryAx} dependency of an equivalent counterpart moves an item from \textit{Statement only} to \textit{Formalized}; unrelated declarations move nothing. Coverage states therefore provide a progress signal that compilable code alone cannot satisfy\rev{. Agent loops such as the propose--judge--update cycle of AlphaApollo~\cite{zhou2025alphaapollo} verify intermediate solutions with tools, which checks the agent's own work but not whether the result is the one the paper states. Coverage states add that check, so} agents act on paper-grounded objectives, and mathematicians see why each task was generated and whether it was completed.

\vspace{-0.1in}

\section{Limitations and Conclusion}
\label{sec:conclusion}

\para{Limitations.}\looseness=-1
Our evaluation covers three projects and 98 items, with few examples in some states, and the ground truth was built by joint consensus, so we cannot report inter-annotator agreement. Semantic correspondence remains model-judged, and \rev{its residual errors depend on the model: Gemini~3.8~Flash still reports some \textit{Partial} items as more complete states, while GPT-6~Sol reports some faithful ones as \textit{Partial}. A \textit{Formalized} state should therefore be confirmed by a person before it is relied on. The proof audit tracks only \texttt{sorry}: a proof resting on an axiom declared by the project would still count as complete, although its certificate would name that axiom. The ground truth takes proof status from the same audit, so this stage is not validated independently.} Extraction assumes structured \LaTeX{} sources. \rev{\textit{LeanTrail} also treats the paper as a fixed specification; when formalization uncovers new results or requirements, as in software verification, declarations that no paper item covers could instead propose revisions to the paper or Blueprint.}

\para{Conclusion.}\looseness=-1
\textit{LeanTrail} treats the source paper as the specification a formalization must meet, keeps proof completeness out of the model's hands, and recovers the correct coverage state for \rev{91.8--92.5\% of items across two judge models and repeated runs}. Re-audited as the repository changes, uncovered results become paper-grounded tasks for formalization agents: progress means the mathematics is complete, not that Lean code compiles.

\vspace{-0.1in}

\begin{ack}
This work was financed in part by the Brazilian National Council for Scientific and Technological Development (CNPq), grant \# 313454/2026-4; the Coordination for the Improvement of Higher Education Personnel (CAPES), Finance Code 001; and the Getulio Vargas Foundation (FGV). Silva was supported by NSF grants OAC-2411221 and DRL-2604563, and by the DARPA expMath program (Agreement No. HR0011262E029). Any opinions, findings, conclusions, or recommendations expressed in this material are those of the authors and do not necessarily reflect the views of NSF, DARPA, or the authors' affiliated institutions. The authors declare no competing interests. 
\end{ack}

% \begin{ack}
% Use unnumbered first level headings for the acknowledgments. All acknowledgments

% Do {\bf not} include this section in the anonymized submission, only in the final paper. You can use the \texttt{ack} environment provided in the style file to automatically hide this section in the anonymized submission.
% \end{ack}

\bibliographystyle{plainnat}
\bibliography{references} 

%%%%%%%%%%%%%%%%%%%%%%%%%%%%%%%%%%%%%%%%%%%%%%%%%%%%%%%%%%%%

\newpage
\appendix
\section{Interface: Paper View and Additional Views}
\label{app:views}

\begin{figure}[H]
  \centering
  \includegraphics[width=\textwidth]{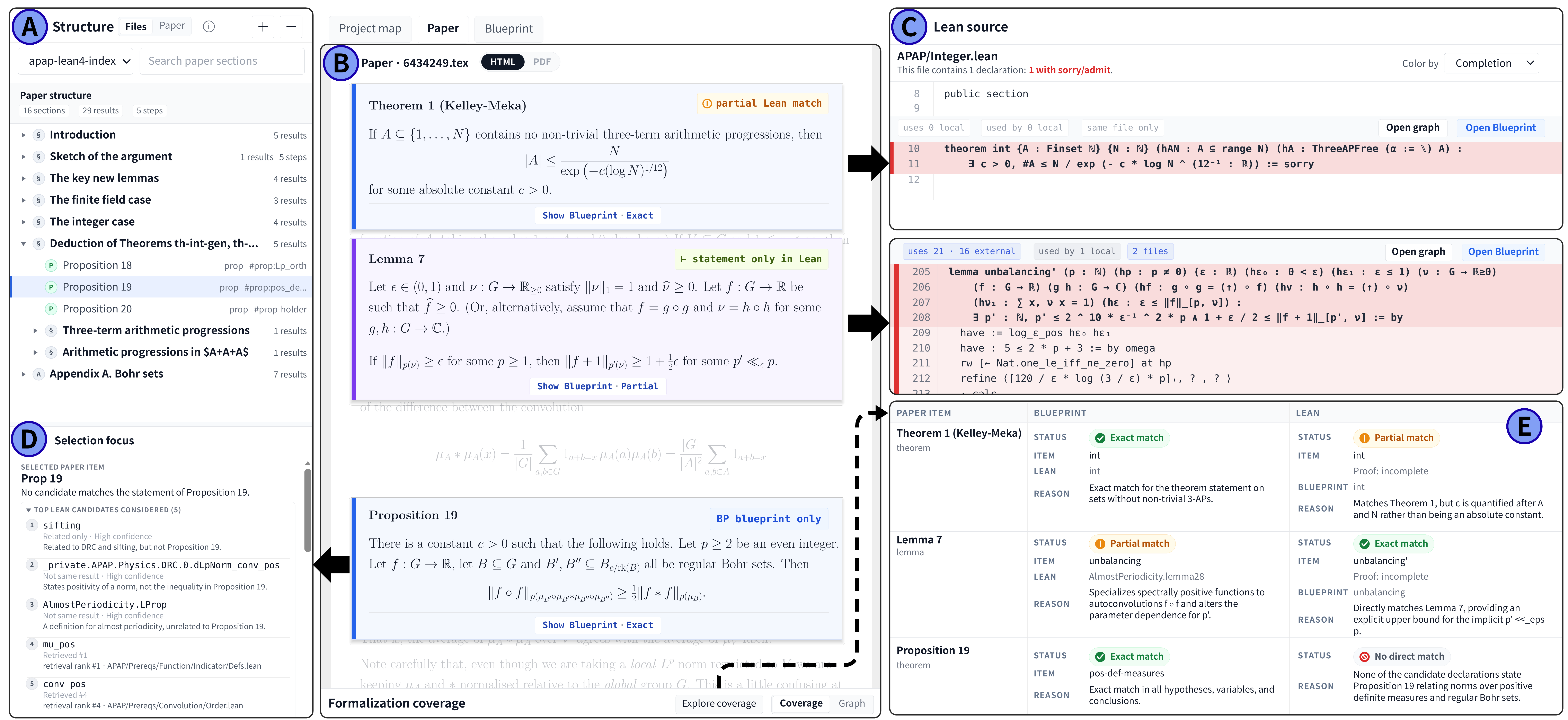}
  \caption{The paper view on the APAP formalization. Extracted paper items are listed in the navigator (\textbf{A}) and inspected in the context of the paper (\textbf{B}). The Lean panels show the selected declarations and their proof status (\textbf{C}); the focus panel exposes candidate judgments with relations and rationales (\textbf{D}); and the coverage table summarizes paper items against Blueprint nodes and Lean declarations (\textbf{E}). Overlaid arrows link each paper item to its Lean counterpart and to its row in the coverage table.}
  \label{fig:paper-alignment}
\end{figure}

The paper view described in \S\ref{sec:inspecting} is shown in Figure~\ref{fig:paper-alignment}. It is one of three coordinated views; the other two are repository-centered and support orientation rather than coverage inspection.

\begin{figure}[H]
  \centering
  \includegraphics[width=\textwidth]{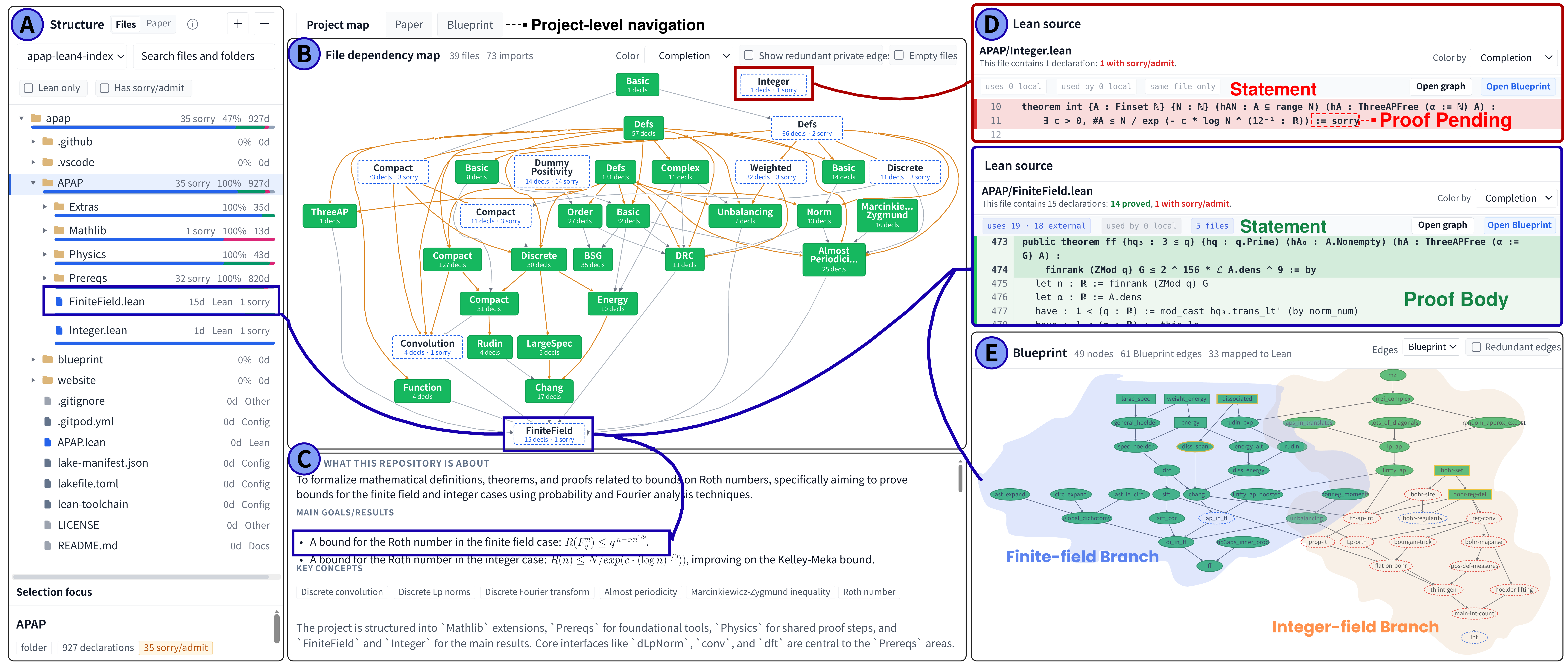}
  \caption{The \textit{LeanTrail} interface on the APAP project, focusing on the project-map and Blueprint tabs. The structural navigator (\textbf{A}), project map (\textbf{B}), repository summary (\textbf{C}), Lean source panel (\textbf{D}), and Blueprint view (\textbf{E}) operate as coordinated views over repository structure, formal code, and the formalization plan. In the project map (\textbf{B}), orange edges denote \texttt{public import} (re-exported dependencies) and gray edges ordinary \texttt{import}; arrows point from the imported dependency to the importing file. Node colors encode proof completion.}
  \label{fig:repo-structure}
\end{figure}

\paragraph{Structural navigator.} The navigator (Figure~\ref{fig:repo-structure}A) operates in two modes. In \textit{Files} mode, users browse the repository's directories and files. In \textit{Paper} mode, it lists items extracted from the paper, so exploration can begin from the mathematical text without knowing where each result lives in the code. Search locates files, declarations, or paper items by name or text fragment, and selections propagate to the other views.

\paragraph{Project map.} The project map (Figure~\ref{fig:repo-structure}B) combines the file-import graph with a repository summary (Figure~\ref{fig:repo-structure}C). The graph exposes central modules, peripheral regions, and files that concentrate incomplete work. The summary complements this structure with the repository's main concepts, thematic areas, and possible entry points for reading. Selecting a node reveals its dependencies and opens the corresponding source.

\paragraph{Blueprint view.} The Blueprint view (Figure~\ref{fig:repo-structure}E) presents the formalization plan, when available, as a navigable artifact. Users can inspect plan nodes, statements, dependencies, and their relations to paper items and Lean declarations. Because the Blueprint lies between paper and code, this view distinguishes nodes already linked to code from those that remain planned or whose correspondence is uncertain.

\paragraph{Code panel and coordinated selection.} The code panel anchors exploration in Lean source: when a selected element has a Lean match, the panel opens its file and highlights the declaration together with its proof-audit status. All views share a common selection mechanism, so a user can start from the map, a paper item, or a Blueprint node and arrive at the corresponding source range.

%%%%%%%%%%%%%%%%%%%%%%%%%%%%%%%%%%%%%%%%%%%%%%%%%%%%%%%%%%%%

\section{Pipeline and Data Model Details}
\label{app:pipeline}

The interface consumes precomputed JSON artifacts rather than inferring alignments interactively, so the coverage table, candidate evidence, proof status, and source panels all read from the same representation. In addition to the entities described in \S\ref{sec:alignment}, extraction computes the declaration-reference and file-import graphs, per-file proof-audit summaries, and source-navigation indices.

\paragraph{Cards and declaration roles.} Paper items, Blueprint nodes, and Lean declarations are normalized into \textit{cards} containing prose, notation, concepts, identifiers, kinds, files, and source locations. Proposition-valued declarations receive an explicit role: a \textit{statement definition} such as \texttt{def Foo : Prop := P} represents the proposition $P$ but does not by itself certify a proof, whereas a \textit{proof-producing} declaration or another term whose type is a proposition may certify one when its dependency closure is complete.

\paragraph{Paper-item extraction.} LeanTrail recursively expands \verb|\input| and \verb|\include| files and recognizes both standard and paper-defined mathematical-statement environments declared with \verb|\newtheorem|. It retains each item's kind, label, statement, section, and source location. The extractor can also conservatively promote independently labeled equations or explicit textual consequences, although none entered this evaluation: all 98 evaluated items came from explicit statement environments. Re-running extraction on the evaluated sources reproduced the same 7, 29, and 62 unique identifiers for FinEqs, APAP, and Ringel, respectively; each occurs exactly once in the corresponding ground truth. The inventories were additionally checked against the source papers before annotation.

%%%%%%%%%%%%%%%%%%%%%%%%%%%%%%%%%%%%%%%%%%%%%%%%%%%%%%%%%%%%

\section{Semantic Judgment Configuration}
\label{app:prompts}

\rev{We evaluate the semantic judgment with two models and the same prompt: \texttt{gemini-3.8-flash} with medium thinking, and \texttt{gpt-6-sol} with medium reasoning effort. Both return schema-constrained JSON with no explicit temperature override.} Each paper item produces one Paper$\rightarrow$Lean call and one Paper$\rightarrow$Blueprint call. The Lean route receives the paper item and Lean statement cards; the Blueprint route receives the paper item and Blueprint statement cards. Declaration bodies, proof-completeness fields, Blueprint provenance in the Lean route, and Lean-link evidence in the Blueprint route are rejected by a recursive whitelist check before prompting.

\paragraph{Diagnostic labels.} Candidate judgments use \textit{exact statement}, \textit{representation-equivalent}, \textit{stronger direct instance}, \textit{partial statement}, \textit{supporting only}, \textit{related only}, \textit{not the same result}, or \textit{no reliable match}; the route-level recommendation may additionally be \textit{ambiguous}. For evaluation and fusion, exact, representation-equivalent, and stronger direct instances collapse to \textit{Equivalent}; partial statements collapse to \textit{Partial}; and supporting, related, rejected, absent, or ambiguous matches collapse to \textit{None}. Up to five candidate judgments per route are persisted with confidence labels and rationales.

\paragraph{Decision policy supplied to the model.} The prompt instructs the model to judge statement-level correspondence only and not to force a match. A faithful-family label requires agreement of substantive hypotheses, quantifiers, domains, parameter restrictions, and conclusions. A candidate known to target the same result is \textit{partial statement} if any such agreement fails or remains materially uncertain; in particular, proving $A\rightarrow B$ when the paper states $B$ is partial when $A$ is substantive. \rev{Before a faithful-family label, the model audits the candidate: it names the paper counterpart of every Lean hypothesis and every parameter restriction, and the Lean counterpart of every paper hypothesis and every conclusion; unfolds local predicates whose definitions appear among the candidates and compares their thresholds with the paper; and checks quantifier order, what each constant may depend on, and whether the hypotheses can hold together. A candidate that proves only part of the paper result, such as one direction, one side of a bound, or one case, is \textit{partial statement}.} A different ingredient or sublemma is \textit{supporting only}, not partial. Differences of representation, for example sets versus predicates or bundled versus unbundled parameters, may be representation-equivalent when the mathematical content agrees. The rationale must identify the concrete representation mapping and why it preserves the substantive statement conditions. For a stronger direct instance, it must specify the instantiation that recovers the paper statement without an additional substantive hypothesis. A proposition definition may be semantically equivalent to the paper statement, but whether it has been proved is deliberately deferred to deterministic fusion.

\paragraph{Prompt template.} The template below summarizes the \textit{LeanTrail Alignment Prompt} used in the reported evaluation, including its route-specific headers and shared decision instructions. Exact fixed templates, including the output-format instructions, are supplied as \path{paper_to_lean_prompt.txt} and \path{paper_to_blueprint_prompt.txt}; \path{response_schema.json} contains the API schema. Each template's final placeholder is replaced by the route-specific JSON packet described below.

\begin{promptpanel}{LeanTrail alignment prompt (summary; verbatim templates in the supplementary material)}
(*@\phead{Paper->Lean header:}@*)

You are judging ONLY the Paper->Lean route for one mathematical paper item.

Question: Does the paper item directly match any local Lean declaration candidate?

- Compare the paper statement to Lean names/signatures only.
- Ignore Blueprint planning evidence in this route.
- A declaration may have entered through a deterministic link, but link provenance is hidden and is not semantic evidence.

(*@\phead{Paper->Blueprint header:}@*)

You are judging ONLY the Paper->Blueprint route for one mathematical paper item.

Question: Does the paper item match any Blueprint node candidate?

- Compare the paper statement to Blueprint titles/statements only.
- Do not judge the linked Lean code in this route.
- A Blueprint->Lean link is metadata for later fusion, not evidence that the Paper->Blueprint alignment is correct.

(*@\phead{Common instructions:}@*)

Classify the statement-level relationship between the paper item and each candidate. Judge mathematical statements only. Proof completeness is unavailable and handled by a deterministic stage.

Do not force a match: no_reliable_match and ambiguous are valid.

(*@\phead{Safety policy:}@*)

- Assign exact_statement, representation_equivalent, or stronger_direct_instance only when all substantive hypotheses, quantifiers, domains, parameter restrictions, and conclusions agree.
- If a candidate clearly targets the same result but any such agreement fails or remains materially uncertain, use partial_statement.
- Use ambiguous when the identity of the target is unresolved.
- Use supporting_only only for a genuinely different ingredient, sublemma, or result.

(*@\phead{Audit before a faithful-family label:}@*)

1. Lean -> paper: name the paper condition matching every Lean hypothesis and parameter restriction;
   an unmatched hypothesis is additional unless it is an instance argument or follows from the setting.
2. Paper -> Lean: name the Lean counterpart of every paper hypothesis and part of the conclusion.
3. Local definitions: unfold project predicates defined among the candidates and compare every
   threshold with the paper; a matching name is not enough.
4. Quantifiers and constants: check quantifier order and what each constant may depend on; a statement
   whose hypotheses cannot hold together, or that is trivially true, does not state the result.
5. The candidate states the paper result itself rather than only a proof ingredient.
If any step fails or remains materially uncertain for a same-target candidate, use partial_statement.

(*@\phead{Definitions:}@*)

- exact_statement: same result and representation.
- representation_equivalent: same mathematical content in a different formal representation.
  It never changes which objects are allowed, a threshold, or the conditions defining a case.
- stronger_direct_instance: genuinely stronger or more general; the paper item follows by direct instantiation without a new substantive argument.
  Use it only when the instantiation can be stated; if it changes what a constant may depend on,
  use partial_statement.
- partial_statement: same intended target, but weakened, narrowed, specialized, incomplete, or subject to an additional substantive hypothesis.
  Proving only one direction, one side of a bound, or one case is partial_statement, not
  no_reliable_match.
- supporting_only: useful for the proof, but not the result itself.
- related_only: shared topic or vocabulary only.
- not_same_result: unrelated or incorrect candidate.
- no_reliable_match: no candidate is reliable enough.

(*@\phead{Additional rules:}@*)

- Type or container differences are not mismatches by themselves.
- Do not call a proof ingredient partial_statement.
- If the paper states B and the candidate proves A -> B for a substantive A absent from the paper, use partial_statement.
- A definition `def Foo : Prop := P` may match P semantically; whether P is proved is decided after semantic judgment.
- Retrieval scores are weak hints only.

Return only schema-valid JSON. Keep at most five candidate judgments and each rationale under 20 words.

For representation_equivalent, use the rationale to identify the concrete representation mapping and why it preserves the substantive statement conditions.

For stronger_direct_instance, use the rationale to specify the instantiation that recovers the paper statement without an additional substantive hypothesis.

(*@\phead{Input:}@*)

<route-specific JSON packet>
\end{promptpanel}

\paragraph{Input packet schema.}
Each judgment call receives a single JSON packet containing the paper item and its retained candidates. The packet below shows the Paper$\rightarrow$Lean form.

\begin{jsonpanel}{Input packet (Paper$\to$Lean route)}
{
  "paper_item_id": "...",
  "display_name": "...",
  "route": "paper_to_lean | paper_to_blueprint",
  "paper_item": {
    "kind": "...",
    "label": "...",
    "title_text": "...",
    "section_path": "...",
    "statement_latex": "..."
  },
  "candidate_bundles": [
    {
      "candidate_id": "...",
      "candidate_source": "lean | blueprint",
      "candidate_type": "...",
      "lean": {
        "decl_id": "...",
        "full_name": "...",
        "kind": "...",
        "declaration_role": "...",
        "file_path": "...",
        "signature": "...",
        "defined_proposition": "..."
      }
    }
  ]
}
\end{jsonpanel}

For the Blueprint route, each candidate bundle replaces the \texttt{lean} object with a \texttt{blueprint} object containing \texttt{blueprint\_id}, label, title, kind, retrieval rank and score, and \texttt{statement\_latex}. Route-inappropriate fields are absent rather than masked with placeholder values.

\paragraph{Response schema.}
The model must return JSON conforming to the schema below, which the API enforces; responses that violate it are rejected and retried.

\begin{jsonpanel}{Schema-constrained model response}
{
  "paper_item_id": "string",
  "final_recommendation": {
    "best_judgment":
      "exact_statement | representation_equivalent |
       stronger_direct_instance | partial_statement |
       supporting_only | no_reliable_match | ambiguous",
    "best_candidate_ids": ["candidate_id"],
    "rationale": "short reason"
  },
  "candidate_judgments": [
    {
      "candidate_source": "lean | blueprint | structural",
      "candidate_id": "candidate_id",
      "judgment":
        "exact_statement | representation_equivalent |
         stronger_direct_instance | partial_statement |
         supporting_only | related_only | not_same_result |
         no_reliable_match",
      "confidence": "high | medium | low",
      "rationale": "short reason"
    }
  ]
}
\end{jsonpanel}

%%%%%%%%%%%%%%%%%%%%%%%%%%%%%%%%%%%%%%%%%%%%%%%%%%%%%%%%%%%%

\section{Annotation Protocol}
\label{app:annotation}

Two annotators familiar with Lean jointly reviewed all 98 items from FinEqs, APAP, and Ringel against the corresponding papers and repository snapshots and, when available, the project Blueprints and author-provided mappings. The evaluated commits are \texttt{c177541} for FinEqs, \texttt{2daecd9} for APAP, and \texttt{7fc71bb} for Ringel. Project-provided links were treated as evidence rather than labels: each proposed match was checked against the paper statement and Lean signature. For each route, the annotators recorded the corresponding candidate identifier or none and the relation \textit{Equivalent}, \textit{Partial}, or \textit{None}; ambiguous and partial decisions include short rationales. All decisions were reached jointly by consensus.

The resulting ground truth contains 7 \textit{Formalized}, 5 \textit{Statement only}, 23 \textit{Partial}, 13 \textit{Blueprint only}, and 50 \textit{Missing} items.

\paragraph{Faithfulness criterion.} \textit{Equivalent} means that the candidate states the same mathematical result, allowing definitional unfolding, notation changes, and mathematically interchangeable representations. A genuinely stronger or more general theorem also counts as equivalent when the paper item follows by direct instantiation without a substantive new argument. \textit{Partial} means that the candidate targets the same result but adds a substantive hypothesis, omits a hypothesis or conclusion, strengthens a parameter restriction, specializes the domain, or otherwise covers only part of the paper statement. A result that is merely useful in a proof, or only shares vocabulary, is \textit{None}.

\paragraph{Coverage-state derivation.} For theorem-like paper items, \textit{Formalized} requires an equivalent local match and a complete proof-producing declaration; an equivalent proposition definition without a proof is \textit{Statement only}. Definition-like paper items may be \textit{Formalized} by a complete implementing declaration. An equivalent selected proof whose audit is incomplete or unknown is \textit{Statement only}. In the absence of an equivalent local Lean match, a partial match on either route yields \textit{Partial}; otherwise an equivalent Blueprint node without a validated local Lean match yields \textit{Blueprint only}. If none of these conditions holds, the item is \textit{Missing}. Proof completeness is derived from the transitive audit and targeted Lean certificates rather than from semantic annotation.

%%%%%%%%%%%%%%%%%%%%%%%%%%%%%%%%%%%%%%%%%%%%%%%%%%%%%%%%%%%%

\section{Evaluation Details}
\label{app:evaluation}

\paragraph{Run provenance.} \rev{Semantic and coverage results come from six evaluation runs of this prompt on all 98 items, three with \texttt{gemini-3.8-flash} and three with \texttt{gpt-6-sol}, each with 196 completed route judgments and no final route errors. The Gemini runs used 198, 199, and 196 provider requests, retrying five invalid structured responses in total; the GPT-6~Sol runs used 196 requests each. Recorded usage was 5.86M prompt and 0.75M output-plus-thinking tokens for Gemini, and 5.42M and 0.36M for GPT-6~Sol.} The same extracted inventories, retrieval candidates, gold annotations, proof evidence, and fusion rules were used for all projects \rev{and runs}. The run and artifact hashes are recorded in the supplementary \texttt{run\_manifest.json}.

\paragraph{Metrics.} Candidate retrieval is evaluated independently on the Paper$\rightarrow$Lean and Paper$\rightarrow$Blueprint routes. A positive query is recovered when at least one annotated counterpart appears among its first $k$ candidates; items whose reference relation is \textit{None} are excluded from this denominator. We report query-level recall@$k$. Semantic judgment is evaluated as three-class classification over \textit{Equivalent}, \textit{Partial}, and \textit{None}, using accuracy and macro-F1. We additionally report whether an annotated target was selected among positive items. Final coverage classification is evaluated using accuracy, macro-F1, and per-state precision, recall, and F1 over the five coverage states. \rev{Semantic and coverage metrics are computed per run and averaged over the three runs of each model.}

\paragraph{Full retrieval results.} \rev{Table~\ref{tab:alignment-results}(a)} reports the complete retrieval curves. At $k=20$, every positive query is recovered on both routes\rev{; at $k=15$, two Paper$\rightarrow$Lean queries are still missed. FinEqs saturates trivially, since its whole Lean inventory has 11 declarations}.

\paragraph{Semantic judgment diagnostics.} \rev{Across all paper items, Paper$\rightarrow$Lean judgment reaches $91.8\pm1.0$\% accuracy and $84.1\pm1.5$\% macro-F1 with Gemini~3.8~Flash and $92.5\pm0.6$\% and $80.7\pm2.0$\% with GPT-6~Sol. Paper$\rightarrow$Blueprint reaches 93.9\% accuracy and 84.7\% macro-F1 with Gemini and 89.8\% and 77.7\% with GPT-6~Sol, with the same accuracy in every run. Among positive items, the models select an annotated target in 30--31/35 (Gemini) and 31--32/35 (GPT-6~Sol) Paper$\rightarrow$Lean cases, and in 35--36/37 (Gemini) and 35/37 in every run (GPT-6~Sol) Paper$\rightarrow$Blueprint cases. On the isolated Paper$\rightarrow$Blueprint route, the \textit{Partial} relation has 55.6\% (Gemini) and 38.5\% (GPT-6~Sol) precision and 71.4\% recall for both.}

\rev{\paragraph{Stability across runs.} The same final state is assigned in all three runs to 96/98 items with Gemini and 97/98 with GPT-6~Sol. The unstable items are APAP's Corollary~9 (\textit{Partial} once, \textit{Missing} twice) and Ringel's Lemma~15 (\textit{Partial} twice, \textit{Formalized} once) under Gemini, and Ringel's Theorem~5 (\textit{Formalized} twice, \textit{Partial} once) under GPT-6~Sol. Seven items are misclassified in every run of each model, so a majority vote over runs does not improve accuracy (90/98 and 91/98).}

\paragraph{Five-state breakdown.} Table~\ref{tab:five-state-breakdown} reports per-state performance for the final coverage classification\rev{, averaged over the three runs of each model. Gemini~3.8~Flash errs in both directions, while GPT-6~Sol recovers every \textit{Partial} item and never reports a more complete state than annotated, at the cost of \textit{Formalized} and \textit{Statement only} recall.}

\begin{table}[H]
\centering
\small
\caption{Per-state performance (\%) of the final coverage classification\rev{, mean over three runs of each model}.}
\label{tab:five-state-breakdown}
\setlength{\tabcolsep}{5pt}
\begin{tabular}{lrrrrrrr}
\toprule
& & \multicolumn{3}{c}{\rev{\textbf{Gemini~3.8~Flash}}} & \multicolumn{3}{c}{\rev{\textbf{GPT-6~Sol}}} \\
\cmidrule(lr){3-5}\cmidrule(lr){6-8}
\textbf{State} & \textbf{$N$} & \textbf{Precision} & \textbf{Recall} & \textbf{F1} & \rev{\textbf{Precision}} & \rev{\textbf{Recall}} & \rev{\textbf{F1}} \\
\midrule
Formalized      &  7 & \rev{82.1} & \rev{85.7} & \rev{83.8} & \rev{100.0} & \rev{52.4} & \rev{68.5} \\
Statement only  &  5 & \rev{50.0} & \rev{60.0} & \rev{54.5} & \rev{100.0} & \rev{20.0} & \rev{33.3} \\
Partial         & 23 & \rev{85.7} & \rev{78.3} & \rev{81.8} & \rev{75.8} & \rev{100.0} & \rev{86.3} \\
Blueprint only  & 13 & \rev{100.0} & \rev{100.0} & \rev{100.0} & \rev{100.0} & \rev{100.0} & \rev{100.0} \\
Missing         & 50 & \rev{98.7} & \rev{100.0} & \rev{99.3} & \rev{100.0} & \rev{100.0} & \rev{100.0} \\
\midrule
Macro           & -- & \rev{83.3} & \rev{84.8} & \rev{83.9} & \rev{95.2} & \rev{74.5} & \rev{77.6} \\
\bottomrule
\end{tabular}
\end{table}

\paragraph{Confusion matrix.} Table~\ref{tab:coverage-confusion} reports the full five-state confusion matrix\rev{, summed over the three runs of each model (294 predictions). Under Gemini, 4 predictions report a \textit{Partial} item as \textit{Formalized} (Ringel's Theorem~10 in every run and Lemma~15 once) and 9 as \textit{Statement only}; the other errors report faithful items as \textit{Partial} or, twice, APAP's Corollary~9 as \textit{Missing}. Under GPT-6~Sol, all 22 errors report a \textit{Formalized} or \textit{Statement only} item as \textit{Partial}.} Coarse final coverage detection (any coverage versus \textit{Missing}) is correct for \rev{292 of 294 Gemini predictions and all 294 GPT-6~Sol predictions}.

\begin{table}[H]
\centering
\small
\caption{Confusion matrix for final five-state coverage classification\rev{, summed over three runs of each model}. Rows are ground-truth states and columns are predicted states.}
\label{tab:coverage-confusion}
\resizebox{\linewidth}{!}{%
\begin{tabular}{lrrrrrrrrrr}
\toprule
& \multicolumn{5}{c}{\rev{\textbf{Gemini~3.8~Flash}}} & \multicolumn{5}{c}{\rev{\textbf{GPT-6~Sol}}} \\
\cmidrule(lr){2-6}\cmidrule(lr){7-11}
\textbf{Gold $\backslash$ Pred.}
& \rev{\textbf{Form.}} & \rev{\textbf{Stmt.}} & \rev{\textbf{Partial}} & \rev{\textbf{BP only}} & \rev{\textbf{Missing}}
& \rev{\textbf{Form.}} & \rev{\textbf{Stmt.}} & \rev{\textbf{Partial}} & \rev{\textbf{BP only}} & \rev{\textbf{Missing}} \\
\midrule
Formalized      & \rev{18} & \rev{0} & \rev{3} & \rev{0} & \rev{0} & \rev{11} & \rev{0} & \rev{10} & \rev{0} & \rev{0} \\
Statement only  & \rev{0} & \rev{9} & \rev{6} & \rev{0} & \rev{0} & \rev{0} & \rev{3} & \rev{12} & \rev{0} & \rev{0} \\
Partial         & \rev{4} & \rev{9} & \rev{54} & \rev{0} & \rev{2} & \rev{0} & \rev{0} & \rev{69} & \rev{0} & \rev{0} \\
Blueprint only  & \rev{0} & \rev{0} & \rev{0} & \rev{39} & \rev{0} & \rev{0} & \rev{0} & \rev{0} & \rev{39} & \rev{0} \\
Missing         & \rev{0} & \rev{0} & \rev{0} & \rev{0} & \rev{150} & \rev{0} & \rev{0} & \rev{0} & \rev{0} & \rev{150} \\
\bottomrule
\end{tabular}%
}
\end{table}

%%%%%%%%%%%%%%%%%%%%%%%%%%%%%%%%%%%%%%%%%%%%%%%%%%%%%%%%%%%%

\section{Agent-Ready Task Export}
\label{app:agent-tasks}

Every non-\textit{Formalized} paper item can be exported as a structured task for a downstream formalization agent. The task type and completion criterion are derived deterministically from the current coverage state, as summarized in Table~\ref{tab:agent-actions}.

\begin{table}[H]
\centering
\small
\caption{Agent actions and completion criteria derived from coverage states.}
\label{tab:agent-actions}
\resizebox{\linewidth}{!}{%
\begin{tabular}{lll}
\toprule
\textbf{Coverage state} & \textbf{Suggested action} & \textbf{Completion criterion} \\
\midrule
Missing
& \texttt{formalize\_missing\_result}
& Validated correspondence is established \\
Blueprint only
& \texttt{implement\_planned\_result}
& Validated local Lean match exists \\
Statement only
& \texttt{complete\_proof}
& Equivalent Lean match with complete proof \\
Partial
& \texttt{repair\_semantic\_alignment}
& Local Lean match becomes Equivalent \\
\bottomrule
\end{tabular}%
}
\end{table}

Each exported task contains the paper item and source location, current coverage state, the retained Lean and Blueprint candidates with their semantic judgments, proof-audit information, the reason the item remains uncovered, and its state-specific completion criterion.

\paragraph{Task schema.}
Each task is serialized as the following JSON object.

\begin{jsonpanel}{Agent-ready task (JSON export)}
{
  "task_id": "...",
  "paper_item_id": "...",
  "coverage_state": "missing | blueprint_only | statement_only | partial",
  "suggested_action": "formalize_missing_result | implement_planned_result | complete_proof | repair_semantic_alignment",

  "paper_item": {
    "display_name": "...",
    "statement_latex": "...",
    "source_file": "...",
    "source_range": "..."
  },

  "lean_candidates": [
    {
      "declaration": "...",
      "relation": "equivalent | partial | none",
      "rationale": "...",
      "proof_status": "complete | incomplete | unknown",
      "source_file": "...",
      "source_range": "..."
    }
  ],

  "blueprint_candidates": [
    {
      "blueprint_id": "...",
      "relation": "equivalent | partial | none",
      "rationale": "..."
    }
  ],

  "reason_uncovered": "...",
  "completion_criterion": "..."
}
\end{jsonpanel}

\paragraph{Re-auditing tasks.} After a new repository snapshot is processed, \textit{LeanTrail} recomputes semantic alignment and proof status against the same paper item and derives its coverage state again. A task is resolved when its completion criterion is satisfied; otherwise, its state and evidence are updated. For example, a \textit{Statement only} task becomes resolved when its equivalent Lean match no longer depends transitively on \texttt{sorryAx}, whereas unrelated additions to the repository leave the task open.

\end{document}